\documentclass[lettersize,journal]{IEEEtran}

\usepackage{algorithm}
\usepackage{stfloats}
\usepackage{url}
\usepackage{verbatim}
\usepackage{cite}
\usepackage{amsmath,amssymb,amsfonts}
\usepackage{algorithmic}
\usepackage[english]{babel}
\usepackage{graphicx}
\usepackage{textcomp}
\usepackage{booktabs}
\usepackage{subfiles}
\usepackage{multirow}
\usepackage{longtable}
\usepackage{mathtools}
\usepackage{gensymb}
\usepackage{array}
\usepackage{booktabs}
\usepackage{placeins}
\usepackage{nomencl}
\usepackage{mdframed}
\usepackage{hyperref}
\usepackage{subcaption}

\newcommand{\Star}[1]{#1\ensuremath{^*}\kern-\scriptspace}
\newcommand{\CStar}{\Star{\ensuremath{\mathrm{}}}}

\IEEEsetsidemargin{i}{1in}

\begin{document}

\title{Nonlinear Model Predictive Control of a Hybrid Thermal Management System}
\author{Demetrius Gulewicz$^{1}$, Uduak Inyang-Udoh$^{2}$, Trevor Bird$^{3}$, and Neera Jain$^{4}$
\thanks{*This work is supported by the U.S. Office of Naval Research Thermal Science and Engineering Program under contract number N00014-21-1-2352.}
\thanks{Demetrius Gulewicz$^{1}$ is a Ph.D. Student in the School of Mechanical Engineering, Purdue University, West Lafayette, IN, USA {\tt\small dgulewic@purdue.edu}}
\thanks{Unduak Inyang-Udoh$^{2}$ is an Assistant Professor in the School of Mechanical Engineering, University of Michigan, Ann Arbor, MI, USA {\tt\small udinyang@umich.edu}}
\thanks{Trevor Bird$^{3}$ is a Senior Lead Engineer at PC Krause and Associates, West Lafayette, IN, USA {\tt\small tbird@pcka.com}}
\thanks{Neera Jain is an Associate Professor in the School of Mechanical Engineering, Purdue University, West Lafayette, IN, USA {\tt\small neerajain@purdue.edu}}
}

\maketitle

\begin{abstract}

Model predictive control has gained popularity for its ability to satisfy constraints and guarantee robustness for certain classes of systems. However, for systems whose dynamics are characterized by a high state dimension,
substantial nonlinearities, and stiffness, suitable methods for online
nonlinear MPC are lacking. One example of such a system is a vehicle
thermal management system (TMS) with integrated thermal energy storage (TES),
also referred to as a hybrid TMS. Here, hybrid refers to the ability to achieve cooling through a conventional heat exchanger or via melting of a phase change material, or both. Given increased electrification in vehicle platforms, more stringent performance specifications are being placed on TMS, in turn requiring more advanced control methods. In this paper, we present the design and real-time implementation of a nonlinear model predictive controller with 77 states on an experimental hybrid TMS testbed. We show how, in spite of high-dimension and stiff dynamics, an explicit integration method can be obtained by linearizing the dynamics at each time step within the MPC horizon. This integration method further allows the first-order gradients to be calculated with minimal additional computational cost. Through simulated and experimental results, we demonstrate the utility of the proposed solution method and the benefits of TES for mitigating highly transient heat loads achieved by actively controlling its charging and discharging behavior.
\end{abstract}


\begin{figure}[!htb]
\centerline{\includegraphics[width=0.95\columnwidth]{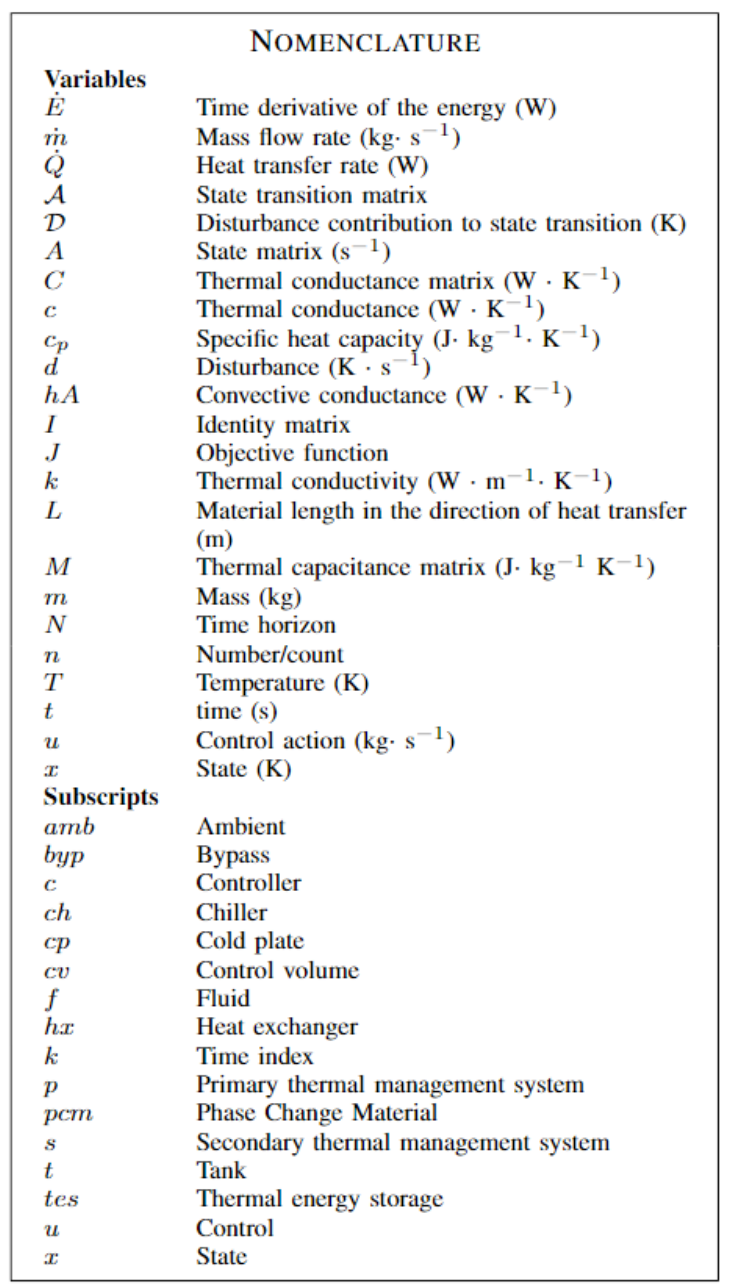}}
\label{fig:nomenclature}
\end{figure}

\section{Introduction} \label{introduction}
\IEEEpubidadjcol
\IEEEPARstart{M}{odel} Predictive Control (MPC) has gained significant popularity as a control methodology for a wide class of systems. There are numerous advantages, including constraint satisfaction and existence of robustness guarantees for certain classes of systems. In general, real-time implementation of MPC involves solving an optimal control problem (OCP) online. An alternative is explicit MPC but this is generally suitable for relatively small problem sizes (1 or 2 controls, short time horizon, up to 10 states) \cite{bemporad_explicit_2019}. As the required computation effort increases, the design of effective numerical methods to formulate and solve the OCP become the principle barrier to the implementation of MPC. To fulfill the real-time requirement, successful MPC applications tend to have at least one of the following characteristics: (1) small problem size relative to the controller update rate, (2) linear or switched linear prediction model, or (3) non-stiff prediction model or timescale separation.
\IEEEpubidadjcol
However, for thermo-fluid systems integrated with latent thermal energy storage (TES), also referred to as hybrid thermal management systems (TMS), none of these characteristics apply.  
One important application of these systems is for thermal management of high-power electronics in vehicle platforms. Such systems can benefit from MPC because of the need for constraint handling, and many researchers have explored MPC for hybrid TMS, albeit in simulation only \cite{candanedo_model-based_2013,shafiei_model_2015,pangborn_hierarchical_2020,leister_nonlinear_2020,vrbanc_simple_2023}; this literature is summarized in Table \ref{tab:lit_review}. Note that for the hierarchical MPC examples, numbers for both the supervisory and lower level controller are reported, with supervisory being reported first. To date, no experimental implementations of nonlinear or linear MPC for a hybrid thermal management system have been published.

\begin{table*}[t]
\begin{center}
\caption{Survey of MPC applied to hybrid thermal management systems}
\label{tab:lit_review}
\begin{tabular}{ c c c c c c c c c}\toprule[1.25pt]
\textbf{Source} & \textbf{TES Model} & \textbf{Controller} & \textbf{States} & \textbf{Steps} & \textbf{Execution Time (s)} & \textbf{Time Step}\\\hline
Candanedo et al. 2013 \cite{candanedo_model-based_2013} & Linear & Single Step & \CStar & 24/30\CStar\CStar & \CStar & 1hr \\
Shafiei and Alleyne 2015 \cite{shafiei_model_2015} & Switched Linear & Hierarchical & 6 & 72/30 & \CStar & 100s/10s\\
Pangborn et al. 2020 \cite{pangborn_hierarchical_2020} & Switched Linear & Hierarchical & 12 & \CStar & \CStar & \CStar\\
Leister and Koeln 2020 \cite{leister_nonlinear_2020} & Nonlinear/Linear & Hierarchical & 3 & 117/10 & 2.2/0.87 & 100s/1s\\
Vrbanc et al. 2023 \cite{vrbanc_simple_2023} & Switched Linear & Single Step & 23 & 24 & 20 & 15 min\\
This Paper & Nonlinear & Single Step & 77 & 25 & 0.5 & 1s\\
\bottomrule[1.25pt]
\end{tabular}
\end{center}
\CStar This information was not available from the cited work. \\
\CStar\CStar An adaptive MPC horizon was used.
\end{table*}

\subsection{Control-oriented Modeling of Hybrid TMS}
For prediction models with a large state dimension and characterized by substantial nonlinearities and stiffness, suitable methods for online nonlinear MPC (NMPC) are lacking. As demonstrated in Table \ref{tab:lit_review}, MPC for hybrid TMS designed using a prediction model with a large number of states is not typical. Importantly, among the papers cited in Table \ref{tab:lit_review}, a computationally tractable controller is achieved by satisfying one or more of the three characteristics mentioned earlier. Under the assumption that the TES device will regularly be allowed to fully solidify or fully melt, simplifying assumptions are reasonable, as demonstrated in \cite{sakakini_switched_2023}. However, for control design that aims to use the TES more flexibly, finer discretization of the PCM volume to dynamically model the melt front is necessary. In our prior work, we showed that modeling the nonlinearities associated with melting and solidification of PCMs often requires a large number of dynamic states \cite{inyang-udoh_model_2023}, especially for real-time estimation of state of charge. In particular, in \cite{shanks_design_2023}, we showed that 21 states were needed for accurate state estimation of a TES device designed as a flat plate heat exchanger with PCM embedded between metal fins, and a PCM volume of approximately $11 \times 15\times 1.3$ cm$^3$. In this paper, we will consider a system with four such devices integrated into a single-phase pumped cooling loop.
 
Another characteristic of many hybrid TMS is that they can exhibit stiff dynamics. Stiffness may arise as a result of large variations in system timescales, and the resulting dynamics can be challenging to solve efficiently \cite{quirynen_numerical_2017}. Additionally, stiffness may arise as a result of the modeling approach. In fact, diffusion equations that replace the second spatial derivatives with finite differences can be stiff \cite[Chapter~IV.1]{hairer_solving_1996}. This is the approach utilized in \cite{shanks_design_2023} to model the TES to obtain the necessary fidelity needed for accurate state estimation. When the TES is integrated into a TMS, the stiffness is exacerbated. In this case, the TMS aims to cool a high-power electronics device whose temperature can change rapidly due to substantial internal heat generation. Together, the higher model order and stiff dynamics are the key characteristics that influence the numerical methods necessary for real-time implementation of MPC designed for hybrid TMS.

\subsection{Integration Methods Suitable for Real-Time Control}
The primary bottleneck for computationally tractable MPC of a hybrid TMS is the integration of the prediction model. Integration of the higher-order dynamics is computationally expensive due to the inherent polynomial time complexity of common matrix operations, especially matrix multiplication and inversion. Stiffness, on the other hand, tends to require less efficient integration routines \cite{moler_7_2004}, which exacerbates the high-order problem by requiring more costly calculations to accurately integrate the dynamics. A general approach to mitigate the high-order problem is to optimize the software and hardware integration of the target embedded system platform. For example, recent developments in basic linear algebra subprograms (BLAS) have reduced computation time for many matrix operations (when compared to existing implementations like MKL and LAPACK), especially for matrix dimensions typical for embedded optimization \cite{frison_blasfeo_2018}. Another way to minimize the computational burden is by optimizing the numerical method. The ideal integrator is maximally efficient at the optimal accuracy for a given MPC. To this end, implicit integration methods have been shown to be particularly efficient for nonlinear stiff systems \cite[Chapter~IV.3]{hairer_solving_1996}. This is in part due to the desirable stability properties that many such methods have, which permit relatively large integration steps in spite of the stiff dynamics \cite{shampine_matlab_1997}. In addition, relatively low integration accuracy is required for many MPC problems, for which single-step Runge-Kutta methods are typically sufficient \cite[Chapter~2]{quirynen_numerical_2017}. Nevertheless, there remain opportunities to exploit characteristics of certain nonlinear dynamical systems such that computational speed can be further improved. 


\subsection{Contribution}
Hybrid thermal management necessitates the synthesis of a nonlinear controller, but selection and implementation of a suitable prediction model, both in terms of model fidelity and numerical integration, has proven difficult. The contribution of this paper is the design and real-time implementation of a nonlinear model predictive controller with a high state dimension (77 states) relative to the controller update rate (1 $s$) on an experimental thermal management system testbed with integrated latent thermal energy storage. We show how in spite of the high dimension stiff dynamics, an explicit integration method can be obtained by linearizing the system dynamics at each time step within the MPC horizon. We then observe that this integration method allows the first-order gradients, commonly used in optimization solvers, to be calculated with minimal additional computational cost. Through simulated and experimental results, we demonstrate not only utility of the proposed solution method but importantly, the benefits of thermal energy storage for mitigating highly transient heat loads achieved by actively controlling the charging and discharging behavior of the TES.

This paper is organized as follows. In Section \ref{System_Model}, the system description and theoretical model for the hybrid TMS are introduced. In Section \ref{sec:NMPC}, the nonlinear model predictive control (NMPC) problem is formulated. In Section \ref{RT_EXP}, the numerical methods used to solve the NMPC problem are presented. In Section \ref{sim_exp_results}, simulated and experimental results are presented, followed by concluding remarks.

\section{System Description} \label{System_Model}

\begin{figure}[htbp]
    \centering
    \begin{subfigure}{0.45\textwidth}
    \includegraphics[width=\linewidth]{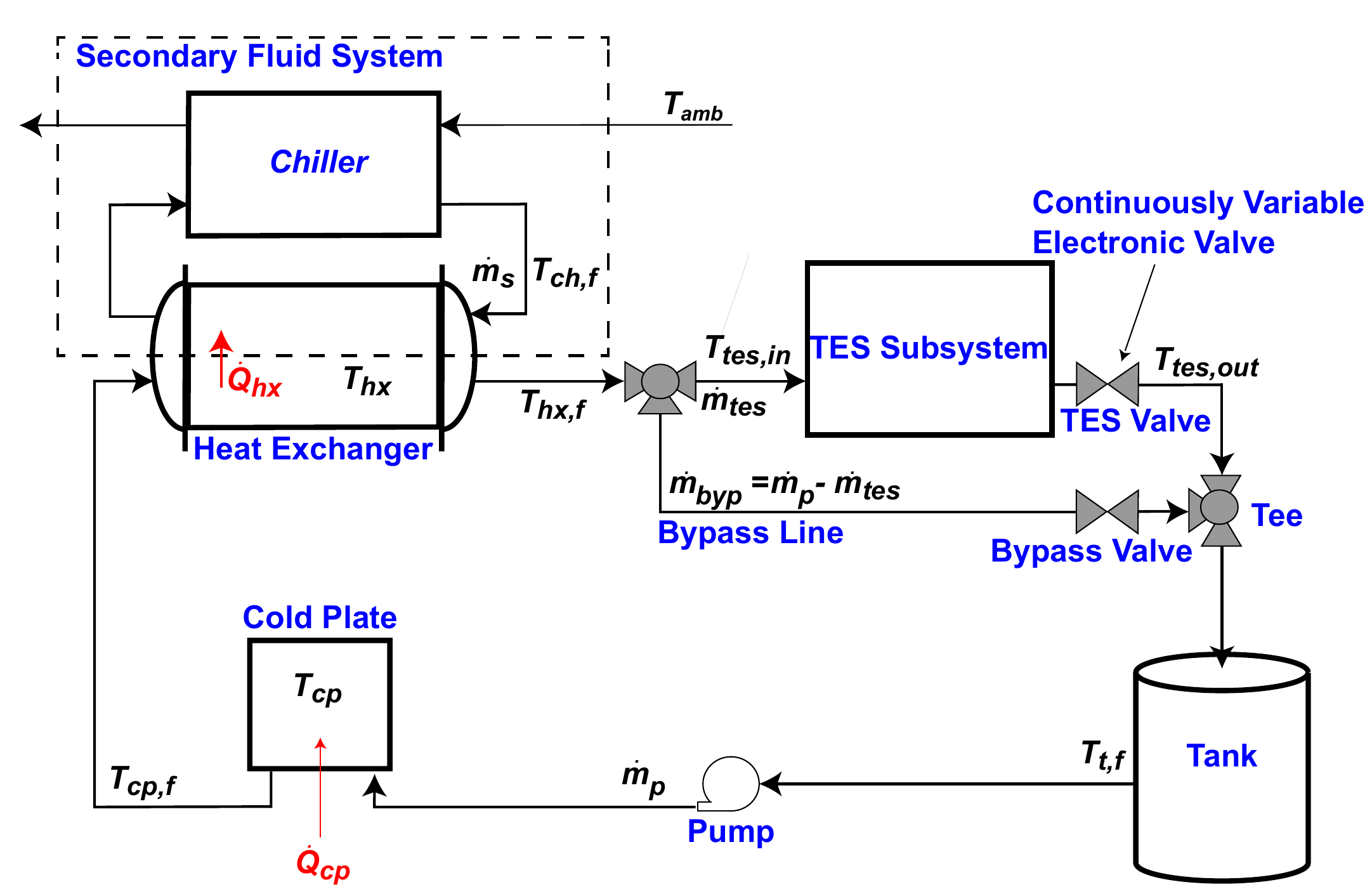}
    \caption{Component-level schematic of the hybrid TMS}
    \label{fig:schematic_1a}
    \end{subfigure}

    \begin{subfigure}{0.45\textwidth}
    \includegraphics[width=0.9\linewidth]{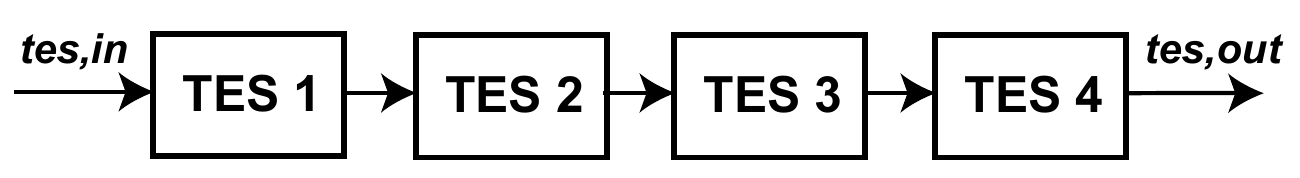}
    \label{fig:schematic_1b}
    \caption{Schematic of the TES subsystem.}
    \end{subfigure}
    
\caption{Schematic of the hybrid TMS.}
\label{fig:schematic_model}
\end{figure}

The hybrid TMS under consideration consists of a fluid reservoir (tank), pump, cold plate, heat exchanger, and latent thermal energy storage (TES) subsystem. A schematic of the system is presented in Fig. \ref{fig:schematic_model}. The pump drives fluid flow through the components with the main objective of absorbing heat from the cold plate and rejecting the heat to a secondary fluid via the heat exchanger and/or temporarily storing excess heat by melting the phase change material within the TES subsystem. Two control valves can be continuously adjusted to regulate the distribution of flow through the TES subsystem and the bypass line. The TES subsystem consists of four identical devices connected in series. Each TES device is a flat plate heat exchanger consisting of finned metal (aluminum) plate separating a phase change material (hexadecane) from the single-phase working fluid (water). The chiller is a vapor compression cycle that exchanges heat with the ambient air ($T_{amb}$), although this system is not modeled. Instead, the temperature $T_{ch,f}$ and mass flow rate $\dot{m}_s$ are assumed to be known disturbances.

\subsection{Hybrid TMS Model}
\subsubsection{Single-phase pumped fluid loop}
We apply a lumped parameter modeling approach to the components that comprise the single-phase pumped fluid loop: the tank, cold plate, and heat exchanger. The temperature of the fluid in the tank is modeled with one state---$T_{t,f}$. The cold plate (CP) and heat exchanger (HX) are modeled with two states each: one for the temperature of the component wall ($T_{cp}$ and $T_{hx}$ respectively), and one for the temperature of the fluid flowing through each component ($T_{cp,f}$ and $T_{hx,f}$ respectively). The governing thermodynamic equations for these components were first presented in \cite{shanks_control_2022} and restated in Eq. \eqref{eqn:non_hybrid_physics}.

\begin{subequations} \label{eqn:non_hybrid_physics}
\begin{align} 
       &(mc_p)_{t,f} \dot{T}_{t,f}={\dot{m}}_p(T_{t,in}-T_{t,f})\\
       &(mc_p)_{cp} \dot{T}_{cp}=(hA)_{cp}\left(T_{cp,f}-T_{cp}\right)+{\dot{m}}_pc_{p}(T_{t,f}-T_{cp})\\
       &(mc_p)_{cp,f} \dot{T}_{cp,f}=(hA)_{cp}\left(T_{cp}-T_{cp,f}\right)+{\dot{Q}}_{cp}\\
       &(mc_p)_{hx} \dot{T}_{hx}=(hA)_{hx}\left(T_{hx,f}-T_{hx}\right)+{\dot{m}}_pc_{p}(T_{cp}-T_{hx})\\
       &(mc_p)_{hx,f} \dot{T}_{hx,f}=(hA)_{hx}\left(T_{hx}-T_{hx,f}\right)+\\&(hA)_{ch}(T_{ch,f} - T_{hx,f}) \nonumber
\end{align}
\end{subequations}

The tank inlet temperature $T_{t,in}$ is given by the convex combination
\begin{equation}
    T_{t,in} = \frac{\dot{m}_{byp}}{\dot{m}_{p}}T_{hx,f} + \frac{\dot{m}_{tes}}{\dot{m}_{p}}T_{tes,out} ,
\end{equation}
where $T_{tes,out}$ is the temperature of the working fluid exiting the final TES device.

\subsubsection{TES Subsystem}
\begin{figure}[!htb]
\centerline{\includegraphics[width=1\columnwidth]{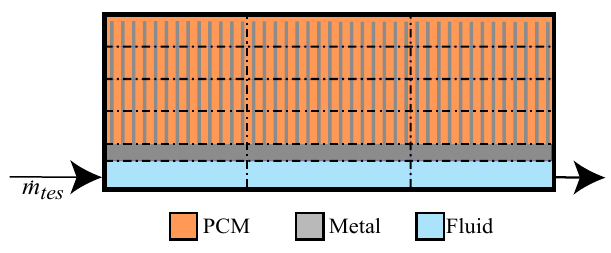}}
\caption{Schematic showing the finite volume discretization of a single TES device into a total of 18 control volumes (3 fluid, 3 metal, and 12 PCM-metal composite).}
\label{fig:TES_subsysystem}
\end{figure}

Given the nonlinearities associated with liquid-solid phase change dynamics, a discretized finite-volume modeling approach is used for each of the thermal energy storage (TES) devices. The reader is referred to  \cite{shanks_design_2023} for a detailed derivation of the model equations, but the modeling assumptions are summarized here for the reader. Each TES device is discretized into a rectangular set of control volumes as depicted in Fig. \ref{fig:TES_subsysystem}.

The fluid channel and metal plate are each discretized using one `layer' of control volumes. Conductive heat transfer in the fluid channel is neglected because it is negligible compared to the advective and convective heat transfer rates. The PCM and fin volume is discretized into a $4 \times 3$ grid of control volumes. The fins are sufficiently thin and closely-spaced such that the material properties of each control volume can be modeled as a composite of the PCM and metal properties \cite{hoe_conductive_2020}. In addition, PCM melting and freezing is modeled via an effective specific heat function. While the specific heat of the liquid and solid phases are different, the effective specific heat of hexadecane during phase change can be approximated as a continuous function \cite{gillis_numerical_2021}. The effective specific heat function enables melt fraction to be computed as a function of the temperatures of each control volume.

For the hybrid TMS model, the state vector $x$ consists of the five non-TES subsystem states (corresponding to the tank, cold plate, and heat exchanger), followed by the $n_{tes}$ TES subsystem states ($T_{tes}$):
\begin{equation}
    {x} = [ T_{t,f},T_{cp},T_{cp,f},T_{hx},T_{hx,f}, T_{{tes}}]^T \in \mathbb{R}^{n_x \times 1} \enspace .
\end{equation}

\noindent The control input $u$ is defined as the mass flow rates through the bypass line and TES subsystem: ${u}=[\dot{m}_{byp},\dot{m}_{tes}]^T \in \mathbb{R}^{n_u \times 1}$. Assuming constant convective heat transfer coefficients, the system is control affine with this particular parametrization. The disturbances consist of the rate of heat transfer to the cold plate (mimicking a head load generated by high power electronics) and the rate of heat transfer from the heat exchanger to the secondary fluid system:
\begin{equation}
    d=[\dot{Q}_{cp},(hA)_{ch}T_{ch,f}]^T \enspace .
\end{equation}

\subsection{Graph-Based State-Dependent Coefficient Representation}
A feature of the governing ODEs of the hybrid TMS are that they can written as a graph of the form
\begin{equation} \label{eqn:state_dependent}
    M(x)\dot{x} = C(x,u,d)x + Bd
\end{equation}
where $M \in \mathbb{R}^{n_x \times n_x}$ is a diagonal matrix comprised of the individual thermal capacitance of each control volume and $C \in \mathbb{R}^{n_x \times n_x}$ is a matrix that describes the thermal resistances between control volumes in the system. Both $M$ and $C$ are state-dependent; this arises from computing the specific heat and thermal conductivity of the PCM, respectively. See the Appendix for a detailed derivation. Finally, $B$ is a constant  matrix that maps the disturbances to the rate of change of energy in each control volume such that $Bd = [0,\dot{Q}_{cp},0,(hA)_{ch}T_{ch}, 0, \cdots 0]^T \in \mathbb{R}^{n_x \times 1}$. Inverting $M$ results in the simplified form of the system model shown in Eq. \eqref{eqn:TMS_simple}. Note that $\tilde{d}$ is not state dependent because the thermal mass of the HX and CP solid control volumes are not state dependent.

\begin{subequations} \label{eqn:TMS_simple}
    \begin{align}
        \dot{x} = A(x,u,d)x + \tilde{d} \enspace \\
        A(x,u,d) = M^{-1}(x)C(x,u,d) \enspace \\
        \tilde{d} = M^{-1}(x)Bd \enspace
    \end{align}
\end{subequations}

Expressing the hybrid TMS model as a graph is helpful because the mathematical properties of graphs can also be exploited for dynamic analysis and control synthesis. For nonlinear MPC, the graph-based formulation also improves computational efficiency of the prediction model by providing a way to compute $C$ with relatively few matrix operations.

\section{Model Predictive Controller Design} \label{sec:NMPC}

To control the hybrid TMS using a nonlinear model predictive controller (NMPC), we consider three principal objectives: 
\begin{enumerate}
    \item High-power electronics device temperature limit (TL)
    \item Thermal endurance (TE)
    \item Power consumption (PC)
\end{enumerate}
The three control objectives are formulated as running costs, with the addition of a penalty on changes in the control action between time steps, and evaluated along the NMPC horizon. The system dynamics are enforced within the computation of the objective function, and the control actions are bounded above and below. These equations are summarized in Eq. \eqref{eq:J}, where $N$ is the number of steps in the NMPC horizon.
\begin{subequations}\label{eq:J}
\begin{align}
\min_{u} J & \triangleq  
\sum_{k=0}^{N-1} \bigg( J_{TL,k+1} + J_{TE,k+1} + J_{PC,k} \\ &\qquad + R_{du}\sum_{j=1}^{n_u}(u_{j,k} - u_{j,k-1})^2\bigg) \notag \\
\text{s.t. } & x_{k+1} = f(x_k,u_k,d_k) \\ \label{eq:sum_constr}
&\sum_{j=1}^{n_u}u_{j,k} \leqslant u_{max} \\ \label{eq:min_constr}
&{u}_{min} \leqslant{u_{j,k}} \\ \label{eq:du_constr}
&|u_{j,k-1} - u_{j,k}| \leqslant \Delta u_{max} \\
&\forall j \in \{1,2\} \nonumber \\
&\forall k \in \{0, \cdots, N-1\} . \nonumber
\end{align}
\end{subequations}
The function $f$ integrates the hybrid TMS dynamics, and will be defined in Section \ref{RT_EXP}. The state dynamics are enforced implicitly by being computed directly in the cost function.

Objective 1 is denoted by $J_{TL}$ and enforces a temperature limit (TL) on the cold plate. While this objective could be enforced as a hard constraint on the state $T_{cp}$, hard constraints can lead to an infeasible optimal control problem (OCP). This can be avoided by introducing soft constraints that heavily penalize violations of the temperature limit in the cost function. Zone control is a typical method, in which additional decision variables are included in such a way that the objective function increases when the relevant states leave the specified `zone' \cite{zuhua_zone_2004}. An alternative option is to use a penalty function directly in the objective function \cite{marvi_safety_2019} which has the benefit of not requiring the addition of states as decision variables. The penalty function, given by

\begin{equation}
\begin{multlined}
J_{TL,k} = \\ \begin{cases}
  \frac{\alpha_1}{T_{cp,max} - T_{cp,k}} + \alpha_2 & T_{cp,k} \leqslant T_{cp,max} - \epsilon \\
  \beta_1 T_{cp,k}^2 + \beta_2 T_{cp,k} + \beta_3 & T_{cp,max} - \epsilon \leqslant T_{cp,k} ,\\
\end{cases} \label{eqn:J_TL}
\end{multlined} \\
\end{equation}
is defined by six scalar parameters: $\alpha_1, \alpha_2, \beta_1, \beta_2, \beta_3, \epsilon$. Four of the six parameters ensure the piece-wise function is twice-differentiable and non-negative for the expected range of temperatures. The parameters $\beta_1$ and $\epsilon$ can be tuned to adjust the shape of the function. The soft constraint temperature boundary is set by $T_{cp,max}$.

Objective 2, denoted by $J_{TE}$, is intended to maximize the thermal endurance (TE) of the system and is defined as
\begin{equation} \label{eqn:J_TES}
J_{TE,k} = Q_{tes}\left( {T}_{tes,k}-T_{ch}\textbf{{1}} \right)^T\left( {T}_{tes,k}-T_{ch}\textbf{{1}} \right) ,
\end{equation}
where $Q_{tes}$ is a positive scalar weighting parameter. In applications where the electronics are safety critical, such as avionics in many air vehicles, thermal endurance refers to the duration of time that a system may operate before cooling objectives cannot be met. Penalizing the temperature states of the control volumes which contain PCM within the TES devices, $T_{tes}$, penalizes candidate solutions that result in less future energy absorption capability. A quadratic expression is used such that the value of t                         he function is zero when maximum energy absorption capability is reached, which is when the TES devices are the same temperature as the chiller temperature. 

Objective 3 is the power consumption (PC) objective, denoted by $J_{PC}$ and defined as
\begin{equation}
J_{PC,k} = R_u \left( \sum_{j=1}^{n_u}u_{j,k}\right)^2 , \\
\end{equation}
where $R_u$ is a positive scalar weighting parameter. The primary mass flow rate is assumed here as a proxy of pumping power and penalized to emphasize solutions that consume less power.




\section{NMPC Solution Method} \label{RT_EXP}

In this section, we describe the numerical methods used to calculate the NMPC solution. First, we show how the system dynamics can be linearized at each time step within the MPC horizon to yield a Runge-Kutta process that has an explicit solution. Then, we observe that the exact solution to these linearized dynamics can also be used to find analytical expressions for the first order gradients often used in optimization solvers.

\subsection{Integration of Simplified Hybrid TMS Dynamics}\label{exp_int}
The hybrid TMS dynamics can be simplified such that each integration step is an explicit function. Consider integrating Eq. \eqref{eqn:TMS_simple} from $t_k$ to $t_{k+1}$ where the time duration of each step in the NMPC horizon is $\Delta t = t_{k+1} - t_{k}$. If $\Delta t$ is chosen to be the update rate of the controller ($\Delta t = t_c$), then $u$ and $\tilde{d}$ are constant from $t_k$ to $t_{k+1}$. The system is linearized by fixing the state-dependent coefficients using the current system state $x_k$. Hence from $t_k$ to $t_{k+1}$, the system dynamics described by Eq. \eqref{eqn:TMS_simple} are linearized as
\begin{subequations} \label{eqn:TMS_linear}
\begin{align}
    \dot{x} = A_k{x} + \tilde{d}_k \enspace \\
    A_k = A(x_k, u_k, d_k)\enspace
\end{align}
\end{subequations}

\noindent for each step in the NMPC horizon. This simplification is well suited for the hybrid TMS under examination because the `fast' cold plate dynamics can be accurately modeled as linear, whereas the nonlinear TES dynamics are much slower. Next, note that the continuous time dynamics can be written in an expanded matrix format, as in Eq. \eqref{eqn:phi_dynamics}.

\begin{subequations} \label{eqn:phi_dynamics}
\begin{align}
    \phi_k = {\begin{bmatrix}
        A_k && \tilde{d}_k \\
        {0} && 0
    \end{bmatrix}} \\
    \tilde{x} = \begin{bmatrix}
        x \\
        1
    \end{bmatrix} \\
    \dot{\tilde{x}} = \phi_k \tilde{x}
\end{align}
\end{subequations}

The family of single-step implicit Runge-Kutta (IRK) methods is a widely accepted choice to integrate stiff systems efficiently \cite{moler_7_2004}. When applied to a linear system, the resultant integration rule reduces to its corresponding stability function, see \cite[Table~3.1]{hairer_solving_1996}. We have found the trapezoidal rule (shown as Eq. \eqref{eqn:trap_nonlin}) is particularly efficient for the integration of the hybrid TMS dynamics. When applied to the simplified hybrid TMS dynamics, the result is an explicit solution for $\tilde{x}_{k+1}$ shown as Eq. \eqref{eqn:trap_lin}.

\begin{equation} \label{eqn:trap_nonlin}
    \Tilde{x}_{k+1} = \Tilde{x}_k + \frac{t_c}{2} \left [f(\Tilde{x}_{k}) + f(\Tilde{x}_{k+1})  \right]
\end{equation}

\begin{equation} \label{eqn:trap_lin}
    \tilde{x}_{k+1} = (I - \frac{t_c}{2}\phi_k)^{-1}(I + \frac{t_c}{2}\phi_k)\tilde{x}_k = D_k^{-1}R_k\tilde{x}_k
\end{equation}

We approximate the matrix inverse $D_k^{-1} = (I - \frac{t_c}{2}\phi_k)^{-1}$ using the Newton-Schulz recursive sequence for approximating matrix inverses \cite{schulz_iterative_1933,pan_improved_1991}. To compute $D_k^{-1}$, Eq. \eqref{eqn:newton_schulz} can be iteratively applied where $D_{k,0}^{-1}$ is the initial guess for $D_{k}^{-1}$. When this method is applied for a fixed number of iterations, the overall integration method consists of a fixed number of matrix operations.

\begin{equation} \label{eqn:newton_schulz}
    D_{k,i+1}^{-1} = D_{k,i}^{-1}(2I - D_kD_{k,i}^{-1}) \enspace
\end{equation}

A single step of the proposed integration routine from $t_k$ to $t_{k+1}$ is enumerated in Algorithm \ref{alg:EXP}. At the first step within the NMPC horizon, the initial guess $D_{-1}^{-1}$ is computed with LU decomposition, noting that $k \in \{0, \cdots, N-1\}$. For all subsequent steps, the previous step inverse is used as the initial guess: $D_{k,0}^{-1} = D_{k-1}^{-1}$.


\begin{algorithm}[H]
\caption{Linear IRK integrator}\label{alg:EXP}
\begin{algorithmic}
\STATE $\mathbf{Input:} \hspace{2mm} x_k \hspace{2mm} A_k \hspace{2mm} \tilde{d}_k \hspace{2mm} D_{k-1}^{-1}$
\STATE $\mathbf{Output:} \hspace{2mm} x_{k+1}$
\STATE $\mathbf{Parameters:} \hspace{2mm} r\hspace{2mm} t_c$

\STATE $x_{k+1} = f(x_k, A_k, \tilde{d}_k, D_{k-1}^{-1}):$ \\[2mm]
\STATE \hspace{0.5cm} $Z_{k} \gets \frac{t_c}{2}\begin{bmatrix}
        A_k && \tilde{d}_k \\
        0 && 0
    \end{bmatrix} $
\STATE \hspace{0.5cm} $D_{k,0}^{-1} \gets D_{k-1}^{-1}$ \\[2mm]

\STATE \hspace{0.5cm} $For \hspace{1mm} i \in \{0, \cdots r\}:$ \\
\STATE \hspace{0.9cm} $D_{k,i+1}^{-1} \gets D_{k,i}^{-1}(2I_{n_s} - (I - Z_k)D_{k,i}^{-1})$ \\[2mm]

\STATE \hspace{0.5cm} $D_{k}^{-1} \gets D_{k,r}^{-1}$ \\[2mm]

\STATE \hspace{0.5cm} $\begin{bmatrix}
        \mathcal{A}_k && \mathcal{D}_k \\
        0 && 1
    \end{bmatrix} \gets D_{k}^{-1}(I + Z_k) $ \\[2mm]
\STATE \hspace{0.5cm} ${x}_{k+1} \gets \mathcal{A}_k{x}_k + \mathcal{D}_k$
\end{algorithmic} \label{alg:exp_int}
\end{algorithm}

\emph{Remark 1.} The stability functions of IRK methods are in fact a subset of rational approximations to the matrix exponential, and are widely used in algorithms to compute the matrix exponential. However, it is common to add an additional scaling step to improve accuracy, which we do not do here \cite{al-mohy_new_2010}. There are several applications that utilize the matrix exponential; see \cite{higham_catalogue_2016} for some examples. Some of these applications may benefit from a faster, albeit less precise matrix exponential computation. More fundamentally, applications that require the computation of the inverse of a matrix, or in general the solution to a system of equations may benefit from the Newton-Schulz method.


\subsection{Approximate Gradients}

The calculation time and error for gradient-based optimization algorithms can be greatly reduced if the gradient of the cost function $J$ with respect to the decision variables $U$ is provided analytically rather than approximated numerically with finite differences. Here, $U$ consists of the concatenation of all $N$ control actions: $U = \big[u_0^T \cdots u_{N-1}^T \big]^T \in \mathbb{R}^{Nn_u \times 1}$. The gradient of $J$ with respect to $U$ is then given by

\begin{equation}
    \frac{dJ}{dU} = \frac{\partial J}{\partial U} + \frac{\partial J}{\partial x_N} \frac{\partial x_N}{\partial U} + \cdots + \frac{\partial J}{\partial x_1} \frac{\partial x_1}{\partial U} ,
\end{equation}
which relies on additional partial derivatives of the states with respect to the inputs and states at previous time steps.



Under the assumption that the state transition matrix $\mathcal{A}_k$ is constant in Algorithm \ref{alg:EXP}, the gradient $x_{k+1}$ with respect to $x_{k}$ is the state transition matrix itself:

\begin{equation} \label{eqn:x_deriv}
    \frac{\partial {x_{k+1}}}{\partial {x}_{k}} \approx \mathcal{A}_{k} \enspace .
\end{equation}
However, the same assumption cannot be made to compute the derivative of $x_{k+1}$ with respect to $u_k$ because it would yield the trivial result that the derivative is the zero matrix. In general, the target gradient for $u_{j,k}$ is

\begin{equation}\label{eqn:grad_u}
    \frac{\partial x_{k+1}}{u_{j,k}} = \frac{\partial \mathcal{A}_k}{\partial u_{j,k}}x_k .
\end{equation}

\noindent Instead, we draw upon the fact that the exact solution to the linearized hybrid TMS dynamics uses the matrix exponential, for which exact expressions to compute Eq. \eqref{eqn:grad_u} have been previously derived \cite{najfeld_derivatives_1995}. We take the same approach as Magee et al. \cite{magee_random-effects_2024} and approximate the derivative by the first term of the power series representation:
\begin{equation} \label{eqn:N0_EQN}
    \frac{\partial \mathcal{A}_k}{\partial u_{j,k}} \approx t_se^{t_sA_k}G_{j,k} ,
\end{equation}
where $G_{j,k} = \frac{\partial A_k}{\partial u_{j,k}}$.

Eq. \eqref{eqn:grad_u} can now be simplified to Eq. \eqref{eqn:grad_approx}, where $E_k = [G_{1,k},G_{2,k},\cdots G_{n_u,k}]$ and $\otimes$ denotes the Kronecker product.

\begin{equation} \label{eqn:grad_approx}
    \frac{\partial{{x}_{k+1}}}{\partial {u}_{k}} \approx t_c\mathcal{A}_kE_k\left[I_{n_u} \otimes {x}_k\right]
\end{equation}

\emph{Remark 2.} Eq. \eqref{eqn:grad_approx} is an approximation of the system gradients with respect to the LTI system presented in Eq. \eqref{eqn:TMS_linear}, which itself is a linearization of the hybrid TMS dynamics. Although it is possible to obtain an analytical gradient using the explicit solution presented in Eq. \eqref{eqn:trap_lin}, we have found Eq. \eqref{eqn:grad_approx} to be sufficiently accurate, consistent with existing literature \cite{magee_random-effects_2024}.

\section{NMPC Implementation} \label{sim_exp_results}

Here we first present the NMPC implemented in simulation on the hybrid TMS model. This is followed by experimental validation of the controller.  For both the simulated and experimental results, the MATLAB function \texttt{fmincon} is used to solve the optimal control problem with the parameters specified in Table \ref{tab:fmincon_opts}; the NMPC parameters are listed in Table \ref{tab:mpc_parameters_1}.

\begin{table}[!htbp]
\centering
\caption{Selected parameter values for the optimizer}
\label{tab:fmincon_opts} 
    \begin{tabular}{ccc}\toprule[1.25pt]
        \textbf{Option} & \textbf{Selected Value} \\
        \hline
        Algorithm & sqp \\ 
        Specify Objective Gradient & True \\
        Constraint Tolerance & 0.002\\
        Optimality Tolerance & 0.005\\
        Step Tolerance & 0.001\\
        Scale Problem & True\\
        Typical X & 0.05\\
         \bottomrule[1.25pt]
    \end{tabular}
\end{table}

\begin{table}[!htbp]
\centering
\caption{Selected parameter values for the NMPC}
\label{tab:mpc_parameters_1} 
    \begin{tabular}{ccc}\toprule[1.25pt]
        \textbf{Parameter} & \textbf{Selected Value} & \textbf{Unit} \\
        \hline
        $N$ & 25 & steps \\ 
        $\Delta t$ & 1 & s  \\
        $u_{min}$ & 0.005 & kg $\cdot$ s$^{-1}$ \\
        $u_{max}$ & 0.1 & kg $\cdot$ s$^{-1}$ \\
        $\Delta u_{max}$ & 0.02 & kg $\cdot$ s$^{-1}$ \\
        $\epsilon$ & 0.3 & $\degree$C \\
        $T_{cp,max}$ & 45 & $\degree$C \\
        $R_u$ & 0.5 & -- \\
        $R_{du}$ & 0.25 & -- \\
        $\alpha_1$ & 0.027 & -- \\
        $\alpha_2$ & 6.75$\times$10$^{-4}$ & -- \\
        $\beta_1$ & 1 & -- \\
        $\beta_2$ & -79.1 & -- \\
        $\beta_3$ & 1564 & -- \\
        $Q_{tes}$ & 2.5$\times$10$^{-6}$ & -- \\
        $r$ & 0 & -- \\
         \bottomrule[1.25pt]
    \end{tabular}
\end{table}

\subsection{Simulated Results} \label{sim_section}

\begin{figure}[!htb]
\centerline{\includegraphics[width=0.9\columnwidth]{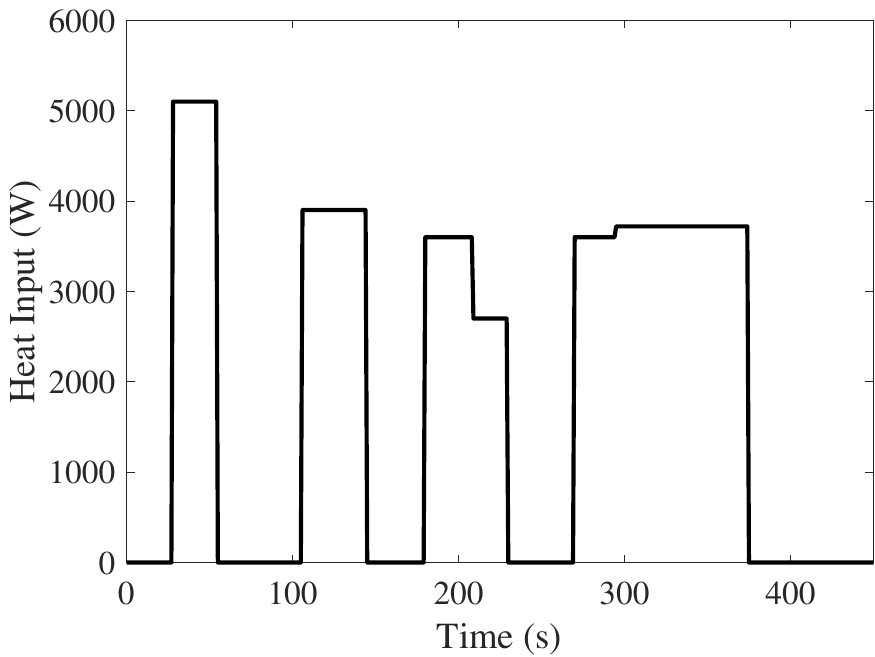}}
\caption{Simulated heat load input applied to the cold plate to emulate high-power electronics.}
\label{fig:heat_load_plain}
\end{figure}

The disturbance profile for the cold plate was designed to test the efficacy of the cold plate soft constraint and to demonstrate significant discharge and recharge of the TES; it is shown in Fig. \ref{fig:heat_load_plain}. In addition, Table \ref{tab:exp_bcs} lists the remaining boundary conditions.

\begin{table}[!htbp]
\centering
\caption{Constant boundary conditions}
\label{tab:exp_bcs} 
    \begin{tabular}{ccc}\toprule[1.25pt]
        \textbf{Boundary Condition} & \textbf{Value} & \textbf{Unit} \\
        \hline
        $\dot{m}_s$ & 0.067 & kg$\cdot$ s$^{-1}$ \\ 
        $T_{ch}$ & 8 & $\degree$C  \\
         \bottomrule[1.25pt]
    \end{tabular}
\end{table}

\begin{figure}[!htb]
\centerline{\includegraphics[width=0.9\columnwidth]{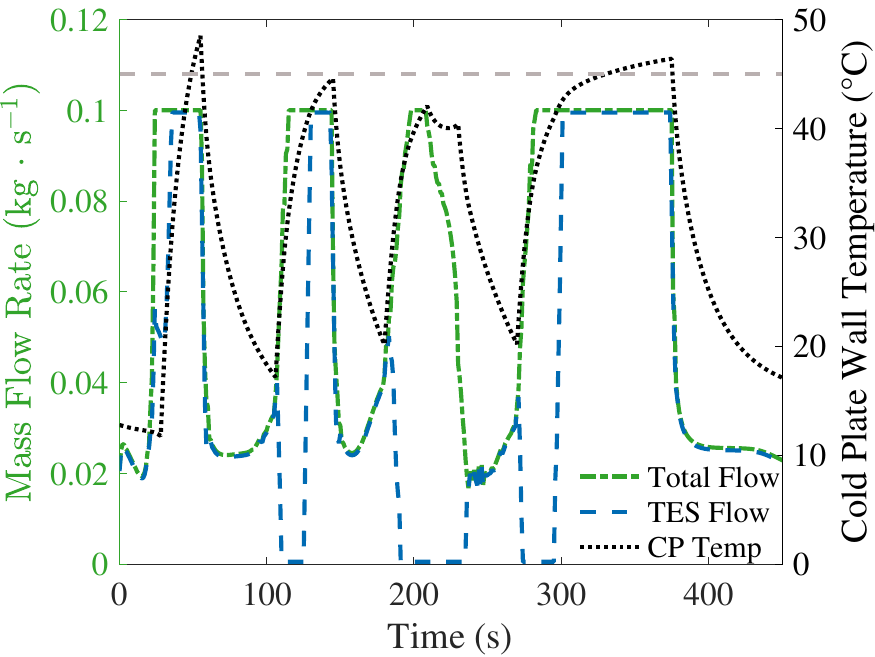}}
\caption{Simulation results depicting the primary (total) mass flow rate through the system, the flow rate through the TES devices, and the cold plate wall temperature.}
\label{fig:sim_dash}
\end{figure}

The NMPC is primarily designed to maintain the temperature of the cold plate below 45 $\degree$C. A plot of the cold plate (wall) temperature (right $y$-axis) and the control inputs (mass flow rates) are shown in Fig. \ref{fig:sim_dash}. The primary fluid mass flow rate, referred to as ``Total Flow'' in Fig. \ref{fig:sim_dash}, increases in anticipation of upcoming heat loads. The first heat load ($t=28$ to $t=55$ seconds) overwhelms the cooling system, so the TES is used maximally to reduce the cold plate temperature. This can be seen from the red dashed curve that denotes the mass flow rate through the TES. The second and third heat load pulses (beginning at $t=106$ and $t=180$ and ending at $t=145$ and $t=230$, respectively) are lower in magnitude than the first, and therefore the cold plate does not exceed the threshold of $45 \degree C$, shown as the grey horizontal line in Fig. \ref{fig:sim_dash}. The TES is used to augment the cooling achieved by the heat exchanger in order to meet the control objectives during each heat pulse. The TES flow rate remains non-zero even when there is no heat load being applied to the cold plate; this not only cools the primary fluid but also enables the TES devices to recharge (i.e. re-solidify). 

\subsection{Experimental Results}

After evaluating the controller in simulation, we implement it on the experimental testbed. The same initial conditions and boundary conditions are used as in the simulation presented in Section \ref{sim_section}.

\subsubsection{Testbed Description}

\begin{figure}[!htb]
\centerline{\includegraphics[width=0.9\columnwidth]{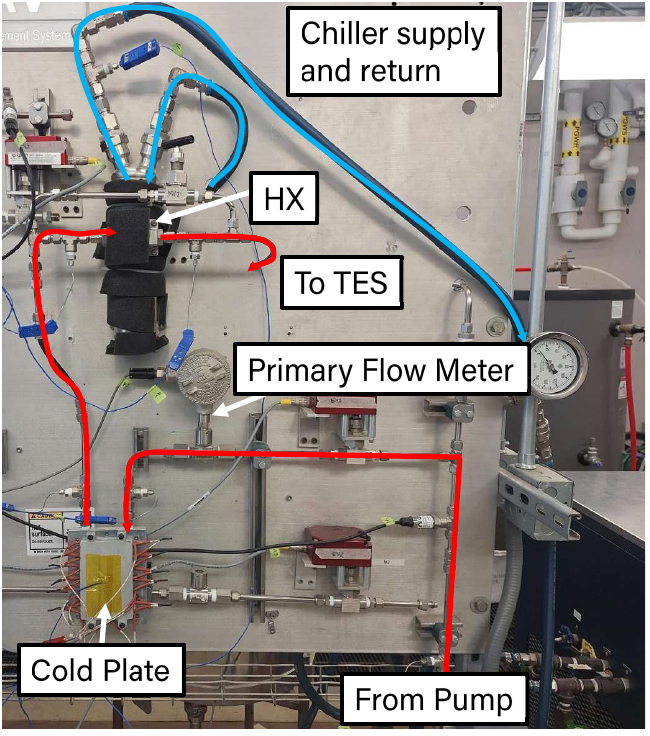}}
\caption{Front side of the hybrid TMS testbed.}
\label{fig:fhTMS}
\end{figure}

The physical testbed is built in the same arrangement as described in Section \ref{System_Model}. A high-power electronics device is simulated by embedding many resistive heaters into an aluminum block and then attaching the block to the face of the cold plate (see Fig. \ref{fig:fhTMS}).

\begin{figure}[!htb]
\centerline{\includegraphics[width=0.9\columnwidth]{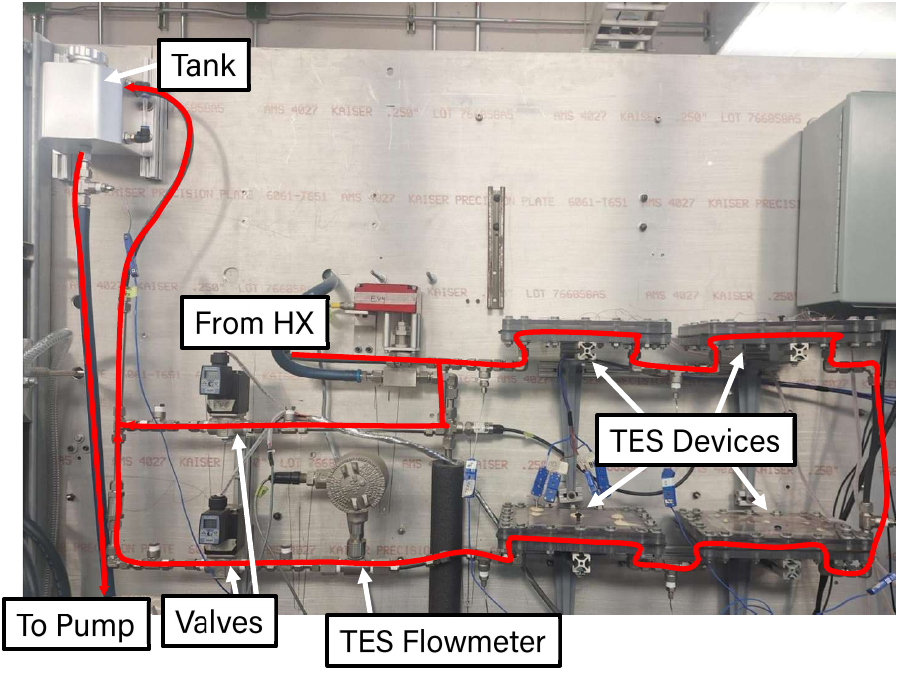}}
\caption{Rear side of the hybrid TMS testbed.}
\label{fig:rhTMS}
\end{figure}

The tank, TES subsystem and bypass line are mounted on the opposite side as shown in Fig. \ref{fig:rhTMS}.

The states for the tank, cold plate, and heat exchanger are directly measured using type T thermocouples. Additionally, twelve of the seventy-two TES states are directly measured with type T thermocouples. The remainder of the states are estimated via a nonlinear observer, described in \cite{shanks_design_2023}.  For each module, the fluid inlet, fluid outlet, and three PCM states are used with the observer to estimate the remaining 15 TES states of that module. A separate observer is used for each of the four modules.

The controller is implemented on a Windows 10 computer with 16GB RAM and an Intel i5-4590 3.3 GHz CPU. A LabVIEW VI calls a MATLAB function which completes the optimization. All other software (including data acquisition, flow rate control and the TES state observers) is implemented on a PXIe 1078 chassis and executes at 10 Hz. An Ethernet connection manages communication between the PXIe and the controller computer.

\subsubsection{Controller Validation}
\begin{figure}[!htb]
\centerline{\includegraphics[width=0.9\columnwidth]{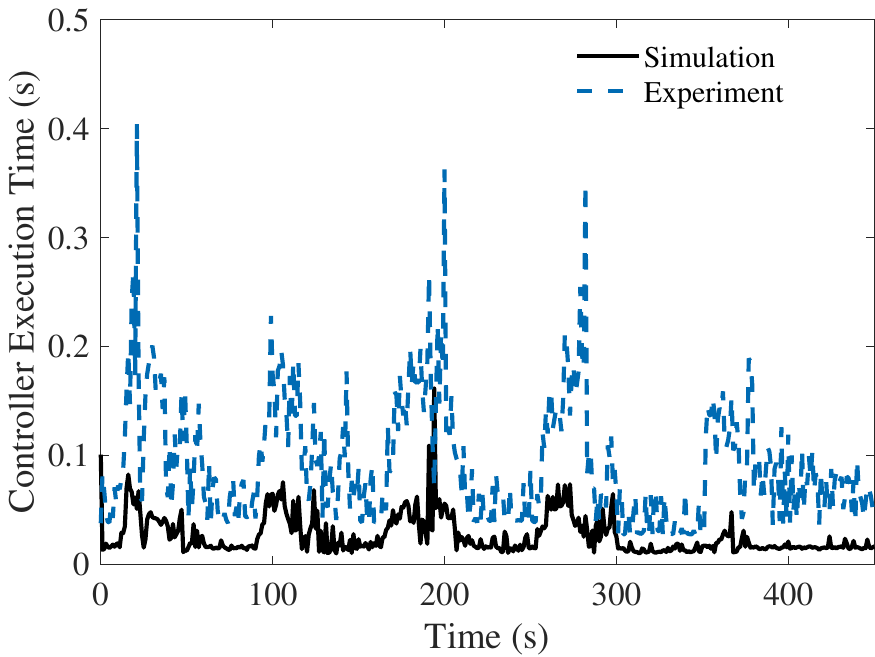}}
\caption{Controller execution time on the experimental testbed.}
\label{fig:exp_time}
\end{figure}

\begin{figure}[!htb]
\centerline{\includegraphics[width=0.9\columnwidth]{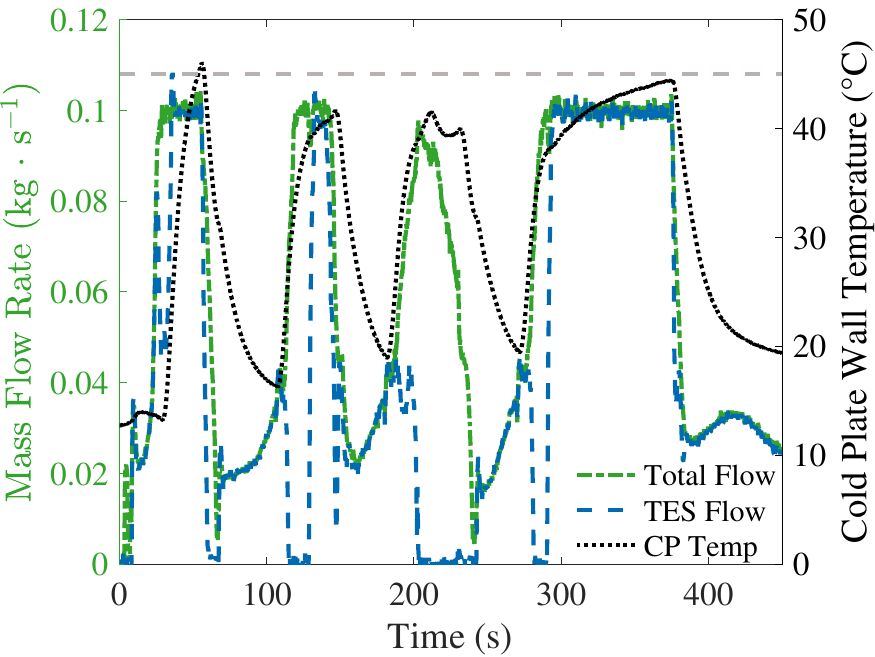}}
\caption{Experimental results: comparison of the total mass flow rate and the flow rate through the TES, plotted against the cold plate wall temperature.}
\label{fig:exp_dash}
\end{figure}

The computation time remains below the controller update rate of 1s. Fig. \ref{fig:exp_time} shows a plot of the controller execution time throughout the duration of the experiment. Fig. \ref{fig:exp_dash} shows the control inputs (mass flow rates) and the resultant cold plate wall temperature. 

\begin{figure}[!htb]
\centerline{\includegraphics[width=0.9\columnwidth]{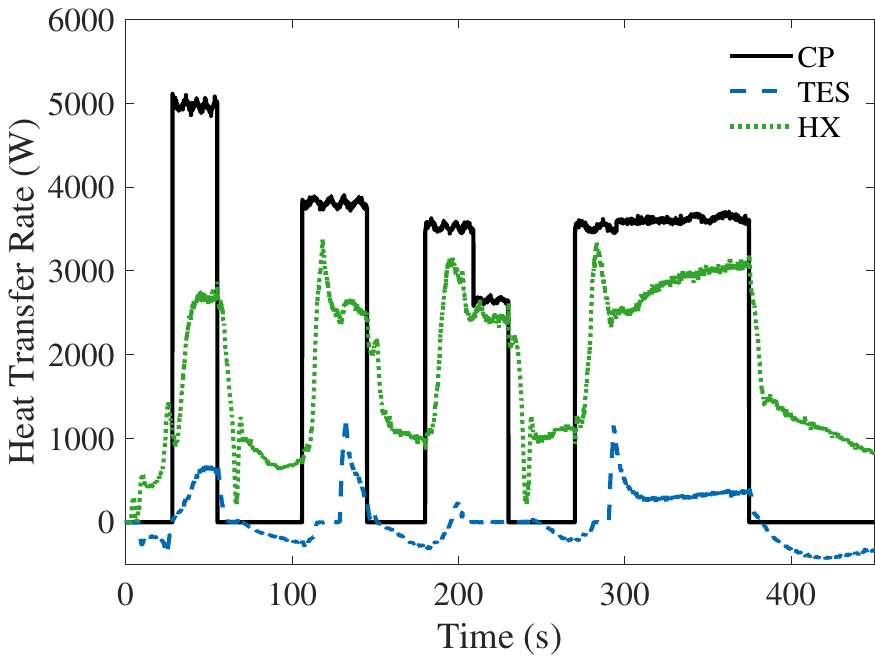}}
\caption{Experimental results: comparison of the heat transfer rates from the primary fluid to the TES and HX, respectively, plotted against the rate of heat transfer applied to the cold plate.}
\label{fig:qdot_exp}
\end{figure}

\begin{figure}[!htb]
\centerline{\includegraphics[width=0.9\columnwidth]{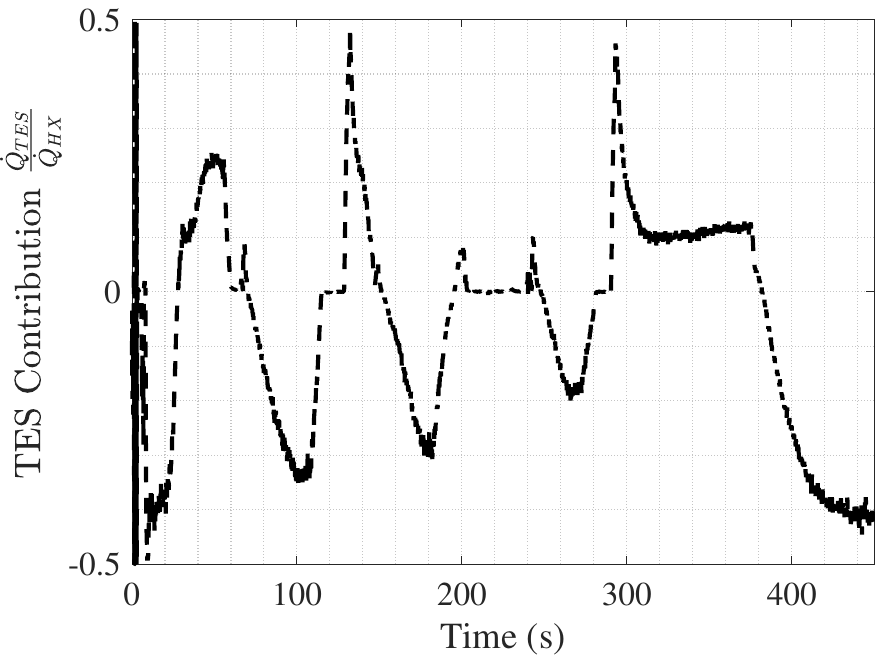}}
\caption{Experimental results: plot of the heat transfer rate from the primary fluid to the TES, relative to the heat transfer rate to the HX.}
\label{fig:tes_contribution}
\end{figure}

A comparison of Fig. \ref{fig:exp_dash} against Fig. \ref{fig:sim_dash} indicates good agreement between the simulated and experimental results.
As expected, during periods of high heat loads, greater utilization of the TES results in a lower cold plate temperature. Another way to interpret the control actions is to examine the rate of heat transfer from the primary fluid to the HX and to the TES. These signals are plotted in Fig. \ref{fig:qdot_exp}, overlaid against the rate of heat addition to the cold plate. Fig. \ref{fig:tes_contribution} shows a plot of the heat transfer rate to the TES relative to the heat transfer rate to the HX. Until the TES flow rate decreases to zero at $t = 200$ seconds (see Fig. \ref{fig:exp_dash}), the TES is recharging (see the negative heat transfer rate plotted in Fig. \ref{fig:qdot_exp}) due to the term $J_{TE}$ in the cost function that is designed to penalize the utilization of the TES. The heat load to the cold plate is sufficiently small that the TES is not needed for additional cooling. Since the heat exchanger is in the primary fluid loop, immediately downstream of the cold plate, it receives the hottest fluid. This configuration was chosen because it maximizes heat removal from the system. However, a consequence of this design is that the temperature of the fluid entering the TES devices is always less than that entering the heat exchanger. This limits the discharge capability of the TES by resulting in a lower temperature differential between the fluid and the PCM. In turn, the TES always contributes less than 50\% of the total heat transfer rate achieved by the hybrid TMS (see Fig. \ref{fig:tes_contribution}).  It is important to note, though, that maximizing utilization of the TES is not the goal of this system; it is primarily to augment the HX as needed to meet the primary control objective of keeping the cold plate wall temperature below the specified threshold. If greater utilization of the TES is desired, tools such as control co-design could be used to design or size the TES and HX accordingly.

\subsubsection{Comparison Against Thermal Management System without TES}

\begin{figure}[!htb]
\centerline{\includegraphics[width=0.9\columnwidth]{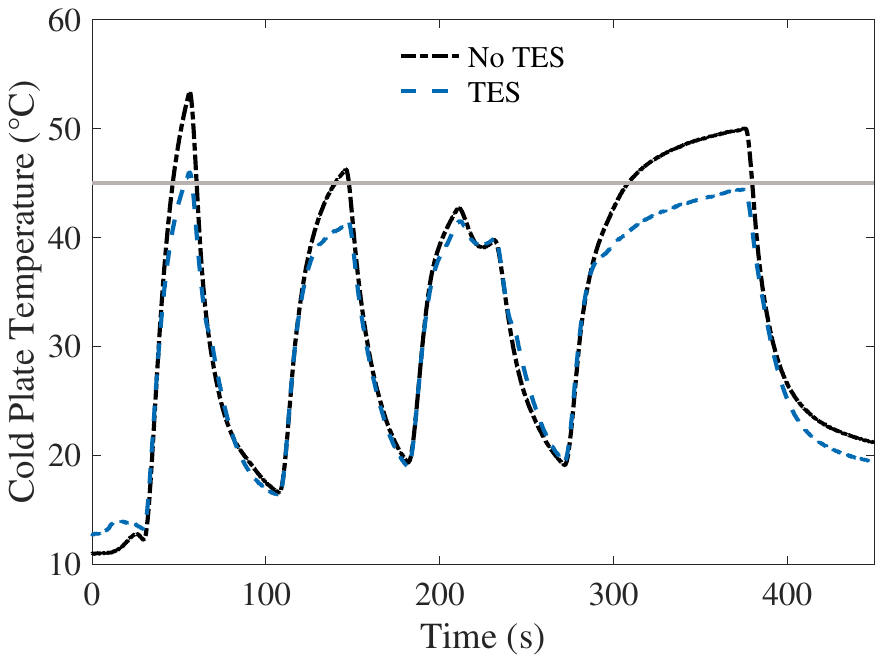}}
\caption{Comparison of the cold plate wall temperature measured during experiments with and without TES.}
\label{fig:tcp_tes_vs_small}
\end{figure}

\begin{figure}[!htb]
\centerline{\includegraphics[width=0.9\columnwidth]{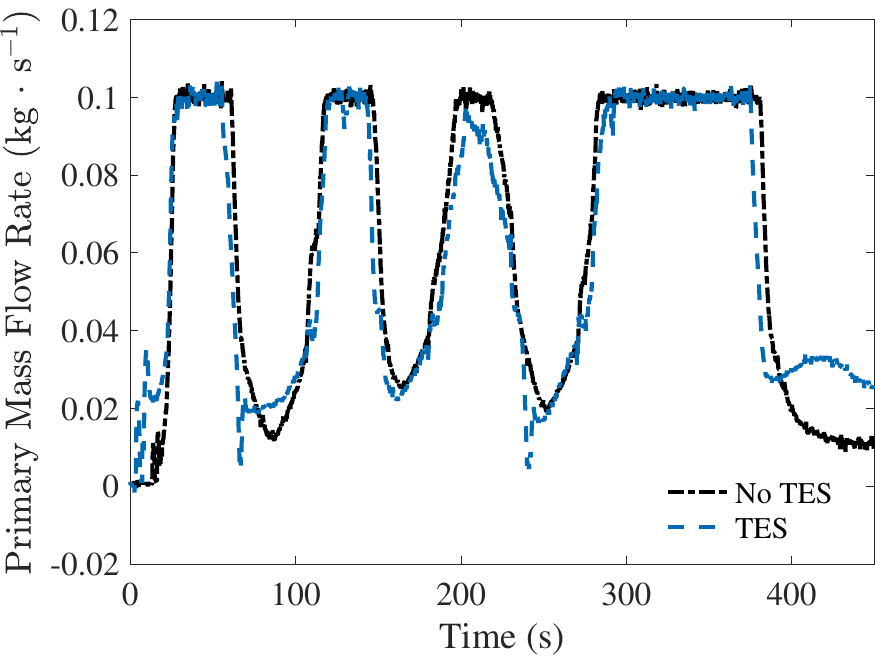}}
\caption{Comparison of the primary flow rate measured during experiments with and without TES.}
\label{fig:mp_tes_vs_small}
\end{figure}

To demonstrate the benefit of integrating thermal energy storage into the thermal management system (TMS), we design and implement an MPC for the TMS only (i.e. without TES). Without the TES, the thermal endurance objective is removed. Fig. \ref{fig:tcp_tes_vs_small} compares the cold plate temperature that is achieved using the hybrid TMS versus the TMS alone, and Fig. \ref{fig:mp_tes_vs_small} compares the primary mass flow rates between the two cases. 

\begin{figure}[!htb]
\centerline{\includegraphics[width=0.9\columnwidth]{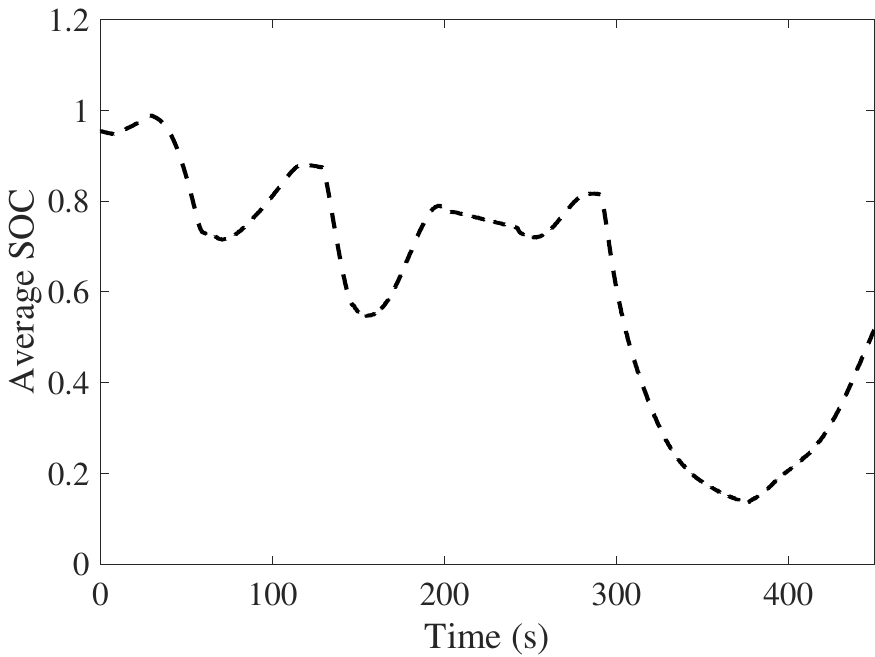}}
\caption{TES state of charge during operation of the hybrid TMS, as estimated using an observer (an exact measurement of the SOC is not available on the testbed).}
\label{fig:tes_soc}
\end{figure}

With the addition of the TES, the cold plate temperature is maintained approximately $7 \degree C$ lower than in the case where there is no TES. Interestingly, the primary flow rates are fairly similar in the two cases, as shown in Fig. \ref{fig:mp_tes_vs_small}. The initial condition for both experiments, $x_0$, is initialized to be nearly equal to the chiller temperature, so there is little benefit from increasing the flow rate until the first heat load is applied at $t=28$ seconds. The most significant difference is at $t = 375$ seconds. The controller demands a larger primary flow rate in the hybrid TMS  (blue dashed curve in Fig. \ref{fig:mp_tes_vs_small}) as compared to the TMS alone (black curve) because doing so enables the TES to recharge.  In other words, in the case of the hybrid TMS, after 400 seconds of operation, the state of charge (SOC) of the TES is close to zero (see Fig. \ref{fig:tes_soc}) so the NMPC for that system wants to use colder fluid to re-solidify the PCM. The other notable difference occurs between $t=180$ and $t=230$ seconds. Since the TES is able to absorb heat from the primary fluid to augment the heat rejection achieved by the heat exchanger, a lower primary flow rate is needed in the hybrid TMS as compared to the TMS alone.

\section{Conclusion}
In this paper we presented the design and real-time implementation of a nonlinear model predictive controller with 77 states on an experimental thermal management system (TMS) testbed with integrated latent thermal energy storage (TES). The reduced-order prediction model is nonetheless high dimension to accurately predict the TES melt front in highly transient operating conditions. The controller approximates the nonlinear dynamics and solves a multi-objective nonlinear model predictive control (NMPC) problem to determine the optimal mass flow rates though the system. We show that an explicit Runge-Kutta integration method specifically tailored for efficient computation can be obtained by linearizing the hybrid TMS dynamics. Furthermore, analytical expressions for the first order gradients of the linearized system can be readily obtained. To ensure the NMPC always finds a feasible solution, a penalty function was designed to maintain the cold plate temperature below a specific threshold in lieu of specifying a hard constraint in the NMPC problem.

After evaluating the controller in simulation, we validated its performance on an experimental testbed. The controller was successfully implemented in real time with a 1 second update rate, and the resultant input and state trajectories matched well against the simulated ones. Importantly, the primary control objective of cooling the cold plate was achieved both in simulation and experimentation. We further compared the performance of the hybrid TMS for mitigating transient loads against a conventional (non-hybridized) TMS (with a comparable NMPC). Especially during large step changes in the heat load, the TES is able to maintain a substantially lower cold plate temperature $T_{cp}$ compared to the same TMS when operated without a TES. This work presents the first experimental demonstration of MPC (linear or nonlinear) on a thermal management system with latent thermal energy storage.  Moreover, the modeling approach and numerical tools used to solve the NMPC online can be applied more broadly to similar classes of systems.

\section*{APPENDIX}
\label{sec:appendix}

Here we describe how the continuous time dynamics of the hybrid TMS can be arranged into the state-dependent form shown in Eq. \eqref{eqn:state_dependent}. The energy balance is given by Eq. \eqref{eqn:E_balance} for each control volume (CV) defined in the hybrid TMS model. 

\begin{subequations} \label{eqn:E_balance}
    \begin{align}
        \dot{E}_{cv} = \dot{E}_{in} - \dot{E}_{out} + \Tilde{d} \\
        \dot{E}_{cv} = M(x)\dot{x} \\
        \dot{E}_{in} - \dot{E}_{out} = C(x,u,d)x
    \end{align}
\end{subequations}

\noindent The rate of energy change of each CV, $\dot{E}_{cv}$, can be decomposed into a diagonal thermal capacitance matrix $M$ and the derivative of the state vector $\dot{x}$ (see Eq.~\eqref{eqn:Ecv}). The thermal capacitance of a CV is $mc_p$, where $m$ and $c_p$ are the mass and specific heat respectively of the CV.
\begin{equation}
    \dot{E}_{cv} =
    \begin{bmatrix}
        (mc_p)_{t,f} & & \\
        & \ddots & \\
        & & (mc_p)_{tes,n_{tes}} \\
    \end{bmatrix}
    \begin{bmatrix}
        \dot{T}_{t,f} \\
        \vdots \\
        \dot{T}_{tes,n_{tes}}
    \end{bmatrix}
    \label{eqn:Ecv}
\end{equation}

\noindent Moreover, the energy moving across the boundary of each CV can be decomposed into three terms:
\begin{enumerate}
    \item energy transfer to the CV at a different temperature than the CV ($\dot{E}_{in}$)
    \item energy transfer from the CV at the CV temperature ($\dot{E}_{out}$)
    \item energy transfer to the CV modeled as a disturbance ($\Tilde{d}$).
\end{enumerate}
This is because except for the disturbance $\Tilde{d}$, only advection and diffusion heat transfer processes are modeled. This yields up to three different types of possible terms for each differential equation, shown in Table \ref{tab:conductance_types}.
\begin{table}[!htbp]
\centering
\caption{Types of heat transfer}
\label{tab:conductance_types}
    \begin{tabular}{ccc}\toprule[1.25pt]
        \textbf{Type} & \textbf{Model} \\
        \hline
        Advection & $\dot{m}c_pT$ \\
        Diffusion & $\frac{kA}{L}T$ \\ 
        Convection & $hAT$ \\
        \bottomrule[1.25pt]
    \end{tabular}
\end{table}

In general the convective heat transfer coefficient $h$, thermal conductivity $k$, and specific heat $c_p$ can be nonlinear functions of temperature and/or fluid flow rate, while the heat transfer area $A$ and material depth $L$ are constant. All three types of heat transfer can be decomposed into a thermal conductance $c$ and a temperature $T$. Note that thermal conductance is defined as the inverse of thermal resistance ($c = R^{-1}$). As such, $\dot{E}_{in}$ and $\dot{E}_{out}$ can be decomposed into a square conductance matrix times the state $x$, as shown in Eq. \eqref{eqn:degree_adjacency}.
\begin{subequations} \label{eqn:degree_adjacency}
    \begin{align}
        \dot{E}_{in} = C_{in}x \\
        \dot{E}_{out} = C_{out}x
    \end{align}
\end{subequations}

The matrix $C_{in}$ has zeros on the diagonal and is non-negative. For energy flowing from CV $j$ to CV $i$, the element in the $ij$ position (using column-row notation) is non-zero and equal to the conductance between the two control volumes. For the hybrid TMS, $C_{in}$ is not symmetric but is substantially sparse given that most control volumes are not thermally connected.

\begin{equation}
    C_{in} = \begin{bmatrix}
        0 & c_{1,2} & \cdots & c_{1,n_x} \\
        c_{2,1} & 0 & \ddots & \vdots\\
        \vdots  & \ddots  & \ddots & c_{n_x-1,n_x}\\
        c_{n_x,1} & \cdots & c_{n_x,2} & 0
         \end{bmatrix}
\end{equation}

\noindent The matrix $C_{out}$ is diagonal because energy transfer from the control volume occurs at the CV temperature and is non-negative.

\begin{equation}
    C_{out} = \begin{bmatrix}
        c_{1,1} & & \\
        & \ddots & \\
        & & c_{n_x,n_x}
         \end{bmatrix}.
\end{equation}

\noindent Except for the entry corresponding to the HX, each diagonal entry in $C_{out}$ is equal to the row sum in the corresponding row in $C_{in}$. For the HX, the disturbance is modeled in a state-dependent form such that the energy flow from the chiller fluid to the HX wall is shown in Eq. \eqref{eqn:sd_HX}.

\begin{equation} \label{eqn:sd_HX}
    Q_{hx,ch} = (hA)_{ch}(T_{ch,f} - T_{hx,f})
\end{equation}

The corresponding convective thermal conductance $hA$ must be added to the diagonal entry of the HX, since the temperature $T_{ch,f}$ is not a state in the model and therefore does not exist in $C_{in}$.

\bibliographystyle{IEEEtran}
\bibliography{IEEE_TCST_2024}

\vfill

\end{document}